\documentclass[adp_arxiv,fleqn]{w-art_arxiv}
\usepackage{times}
\usepackage{w-thm}
\usepackage{latexsym}

\theoremstyle{plain}

\theoremstyle{definition}

\usepackage[]{graphicx}
\chardef\bslash=`\\ 

\begin{document}

\DOIsuffix{theDOIsuffix}
\Volume{12}
\Issue{1}
\Month{01}
\Year{2003}
\pagespan{1}{}


\title[On the interaction of mesoscopic quantum systems with gravity]{On the interaction of mesoscopic quantum systems with gravity}


\author[C. Kiefer]{Claus Kiefer\footnote{Corresponding
     author: e-mail: {\sf kiefer@thp.uni-koeln.de}}} \address[]{Institut f\"ur Theoretische Physik, Universit\"{a}t zu K\"oln, Z\"ulpicher Str.~77, 50937 K\"oln, Germany}
\author[C. Weber]{Carsten Weber}
\begin{abstract}
We review the different aspects of the interaction of mesoscopic quantum systems with gravitational fields. We first discuss briefly the foundations of general relativity and quantum mechanics. Then, we consider the non-relativistic expansions of the Klein-Gordon and Dirac equations in the post-Newtonian approximation. After a short overview of classical gravitational waves, we discuss two proposed interaction mechanisms: (i) the use of quantum fluids as generator and/or detector of gravitational waves in the laboratory, and (ii) the inclusion of gravitomagnetic fields in the study of the properties of rotating superconductors. The foundations of the proposed experiments are explained and evaluated.
\end{abstract}
\maketitle                   



\newcommand{\be}{\begin{equation}}
\newcommand{\ee}{\end{equation}}
\newcommand{\beann}{\begin{eqnarray*}}
\newcommand{\eeann}{\end{eqnarray*}}
\newcommand{\bea}{\begin{eqnarray}}
\newcommand{\eea}{\end{eqnarray}}
\newcommand{\lb}{\label}
\newcommand{\bdm}{\begin{displaymath}}
\newcommand{\edm}{\end{displaymath}}
\newcommand{\D}{{\rm d}}
\newcommand{\E}{{\rm e}}
\newcommand{\I}{{\rm i}}
\newcommand{\X}{{\mathbf x}}
\newcommand{\Y}{{\mathbf y}}
\newcommand{\tab}{\hspace{5mm}}
\newcommand{\bi}{\begin{itemize}}
\newcommand{\ei}{\end{itemize}}
\newcommand{\bc}{\begin{center}}
\newcommand{\ec}{\end{center}}
\newcommand{\ul}{\underline}
\newcommand{\ovl}{\widehat}
\newcommand{\bra}[1]{\langle #1|}
\newcommand{\ket}[1]{|#1\rangle}
\newcommand{\braket}[2]{\langle #1|#2\rangle}
\newcommand{\exVal}[1]{\langle #1 \rangle}
\newcommand{\scProd}[3]{\langle #1|#2|#3 \rangle}
\newcommand{\scProdtoo}[2]{\langle #1|#2 \rangle}
\newcommand{\eps}{\epsilon}

\section{Quantum theory and the gravitational field}

In this introductory section we shall give a brief review of the
relation between quantum theory and gravity. For more details and
references we refer to \cite{OUP}. 

The gravitational interaction is distinguished by the fact that
it interacts {\em universally} with all forms of energy. It dominates on large
scales (relevant for cosmology) and for compact objects
(such as neutron stars and black holes). All presently known features of
gravitational physics are successfully described by the theory of
general relativity (GR), accomplished by Albert Einstein in 
1915.\footnote{Exceptions may be the Pioneer anomaly \cite{Nieto} and the rotation
curves of galaxies \cite{Boerner}, which could in principle demand a more general
theory. This is unclear at the moment.} 
In this theory, gravity is not interpreted as a force acting on
space and in time, but as a representation of the {\em geometry} of spacetime.
This is a consequence of the equivalence principle stating the (local)
equivalence of free fall with the gravity-free case.
Mass generates curvature which in turn acts back on the mass. Curvature
can also exist in vacuum; for example, it can 
propagate with the speed of light in the form of 
{\em gravitational waves}.
In contrast to other physical theories, spacetime in GR plays a
dynamical role and not the role of a rigid background structure.

Experimentally, GR is a very successful theory
\cite{Will}. Particularly impressive
examples are the observations of the double pulsar PSR 1913+16
by which the existence of gravitational waves has been
verified indirectly. Many interferometers on Earth are now in operation
with the goal of direct observations of these waves. This would open
a new window to the universe (`gravitational-wave astronomy').
After 40 years of preparation, the
ambitious satellite project Gravity-Probe~B was launched in April~2004
in order to observe the `Thirring--Lense effect' predicted by GR.
This effect states that a rotating mass 
(here the Earth) forces local inertial systems
in its neighbourhood to rotate with respect to far-away objects. 
GR has also entered everyday life in the form of the global positioning system
GPS; without the implementation of GR, inaccuracies in the order of
kilometres would easily arise over the time span of days \cite{Ashby}. 
A recent survey of GR tests in space can be 
found in \cite{OB} and the references therein.

GR can be defined by the `Einstein--Hilbert action',
\be
S_{\rm EH}= \frac{c^4}{16\pi G}\int_{\mathcal M}{\rm d}^4x\ \sqrt{-g}
 \left(R-2\Lambda\right)\ ,
\lb{EH}
\ee
where $c$, $G$, and $\Lambda$ are the speed of light, the
gravitational constant, and the cosmological constant, respectively;
$g$ denotes the determinant of the four-dimensional metric and
$R$ the Ricci scalar.
The integration is over a four-dimensional manifold
${\mathcal M}$, and a surface term has been omitted for simplicity.
In addition one has an action, $S_{\rm m}$, describing non-gravitational
fields minimally coupled to the gravitational field. 
 From it one can derive the energy--momentum tensor
\be
T_{\mu\nu}=\frac{2}{\sqrt{-g}}\frac{\delta S_{\rm m}}{\delta g^{\mu\nu}}\ ,
\lb{em}
\ee
which acts as a `source' for the gravitational field.
The variation of $S_{\rm EH}+S_{\rm m}$ leads to Einstein's field
equations,
\be
G_{\mu\nu}\equiv R_{\mu\nu}-\frac{1}{2}g_{\mu\nu}R+\Lambda g_{\mu\nu}
= \frac{8\pi G}{c^4}T_{\mu\nu}\ .
\lb{einstein}
\ee
{}From general features (`singularity theorems') one can see that
GR specifies---in contrast to Newtonian gravity\footnote{Note, however, that
Newtonian cosmology leads to intrinsic inconsistencies.}---its own limits:
there are situations where GR can no longer be valid. Typically,  
this happens inside black holes and near the Big Bang. The general belief
is that only the inclusion of quantum theory can describe such situations.

Quantum theory is of a universal nature, too. In contrast to GR, however,
it gives the general frame for theories describing particular interactions.
In fact, all known interactions besides gravity (that is, the strong
and electro-weak interactions) are successfully described by quantum theory.
At the heart of quantum theory is the {\em superposition principle},
which is experimentally well established and which is crucial for many
modern developments such as quantum information theory, cf. \cite{deco}.

A consistent application of the quantum framework to the gravitational
field is referred to as {\em quantum gravity}. The construction of such a
theory is one of the big open problems in theoretical physics.
What are the main motivations for the search of such a theory?
\begin{itemize}
\item All non-gravitational interactions are described by quantum theories.
Since gravity interacts universally with all quantum fields,
it is natural to expect that it is described by a quantum theory as well.
The prevailing idea of {\em unification} of all interactions gives
an argument for the existence of a unified quantum theory including gravity.
\item The origin of the universe and the evaporation of black holes
cannot be understood within the classical theory. Cosmology 
and black-hole physics are thus
incomplete without quantum gravity.
\item Ordinary (non-gravitational) quantum theory and GR contain concepts
which are incompatible: whereas quantum theory uses a
rigid background structure (Newton's absolute time $t$ in quantum mechanics,
or Minkowski spacetime in quantum field theory
or---in situations where this approximation holds---in
relativistic quantum mechanics), spacetime in GR
is dynamical. This is sometimes called the {\em problem of time}.
There should therefore exist an encompassing theory of quantum gravity
in which these concepts are modified. In fact, the straightforward
quantization of GR leads to the result that no external time 
can exist at all at the most fundamental level.
\end{itemize}

At present, there exist two main classes of approaches in 
quantizing gravity \cite{OUP}. 
One is {\em quantum GR}, which is obtained by the application of
quantization rules to GR. Among these, the path-integral method
and canonical quantization are especially noteworthy. Preliminary
results seem to indicate a discrete structure of spacetime at
the fundamental level and an avoidance of the classical singularities.
The alternative to this class is {\em superstring theory} (`M-theory').
In contrast to quantum GR, it aims at a unified quantum theory of
all interactions with gravity. Quantum gravity then follows for 
sufficiently low energies when the gravitational interaction can
be considered as a separate entity. Superstring theory also seems to
indicate a discrete structure of spacetime. It predicts the
presence of effective terms violating the equivalence principle.
This could be subject to experimental tests \cite{CL1}.
In addition, one can consider {\em effective theories} of quantum gravity,
which are valid only for low enough energies. They may be either
derived from a fundamental theory or from heuristic considerations.
 From such effective theories one can derive, for example,
quantum-gravitational correction terms to the Newtonian potential
\cite{OUP}. 

The fundamental scale of quantum gravity is the {\em Planck scale}.
It is obtained by combining the speed of light ($c$), 
the gravitational constant
($G$), and the quantum of action ($\hbar$) into units of length, time, and
mass. They read explicitly
\bea 
l_{\rm P} &=& \sqrt{\frac{\hbar G}{c^3}} \approx 1.62\times 10^{-33}\ 
{\rm cm}\; ,
\lb{lP}\\
t_{\rm P} &=& \frac{l_{\rm P}}{c}=\sqrt{\frac{\hbar G}{c^5}}
\approx 5.40\times 10^{-44}\ {\rm s}\; ,\\
\lb{tP}
m_{\rm P} &=& \frac{\hbar}{l_{\rm P}c}=\sqrt{\frac{\hbar c}{G}}
\approx 2.17\times 10^{-5}\ {\rm g}\approx 1.22 \times 10^{19}\ {\rm GeV}\ .
\lb{mP}
\eea 
It is believed that it is
the smallness (or largeness) of these Planck units that has
hitherto prevented a direct experimental access to test theories of
quantum gravity.

Historically, an attempt to combine ordinary quantum theory with
classical gravity in a consistent way has led to the formulation of the
so-called `semiclassical Einstein equations', in which the 
expectation value of the operator-valued energy--momentum tensor 
with respect to some quantum state is coupled to the 
classical gravitational field.
Instead of (\ref{einstein}) one would then have
\be
\lb{semi}
R_{\mu\nu}-\frac{1}{2}g_{\mu\nu}R+\Lambda g_{\mu\nu}=
  \frac{8\pi G}{c^4}\langle\Psi\vert\hat{T}_{\mu\nu}
\vert\Psi\rangle\ .
\ee
However, it is now generally believed that these equations can at best
be an approximation. First, a `Schr\"odinger cat-type experiment'
performed by Page and Geilker \cite{PG} has given
results in contradiction to (\ref{semi}). These authors have considered
a box in which there are two rigidly connected masses and a
quantum-mechanical system that is in a superposition of `being decayed'
and `being non--decayed'. In case of a decay, a mechanism is put into work
that destroys the rigid connection of the masses and enables them to approach 
each other. The superposition of the quantum-mechanical system is then
transferred to the masses. Outside the box is a Cavendish balance.
According to (\ref{semi}), the balance would follow the
expectation value of the mass distribution and slowly swing in time.
The experiment, however, shows an abrupt deflection occurring at a certain
time, in conflict with (\ref{semi}).

The second point concerns the investigation
of models describing the coupling of a quantum system to a classical
system (`hybrid models'), which seems to lead to inconsistencies. Equation
(\ref{semi}) usually holds if the mean square deviation (`fluctuation') of
$\hat{T}_{\mu\nu}$ is small. This condition is violated in the
Page--Geilker experiment. It is also violated in another interesting
situation---the gravitational radiation emitted by quantum systems, as
discussed by Kuo and Ford \cite{KF}.
There the predictions of a linearized theory of quantum gravity
differ drastically from (\ref{semi}). Another case where the 
fluctuations of the energy--momentum tensor are large occurs when
$\langle\Psi\vert\hat{T}_{\mu\nu}\vert\Psi\rangle$ can become negative.
This happens, for example, in the Casimir effect or the case of a squeezed
vacuum (`particle creation') relevant in cosmology or black-hole physics.\\

This study is devoted to the interaction of mesoscopic quantum systems
(including for definiteness also microscopic quantum systems)
with gravity. The main motivations are the following:
\begin{itemize}
\item As far as the relation of quantum theory with gravity is
concerned, this is the only level where {\em experiments}
have been either performed or have been suggested. A clear understanding of
this interaction would thus be crucial for the development of
quantum gravity. It could, in particular, help to obtain
conceptual clarity. Moreover, approaches to quantum gravity suggest
that there may be effects (such as the 
above-mentioned violation of the equivalence
principle from string theory) that could be observed in this regime.
\item It is of interest to study the realizability of experimental
proposals connected with current theoretical work.
 This could lead to a decision about ideas predicting
the observability of surprising effects. This will be discussed in
detail in Sects.~3 and~4. 
\item This interaction can be used as a {\em tool} \/ for other
experiments. One example is the planned ESA satellite mission HYPER. It will use atom interferometry to test the equivalence principle
and to measure the Thirring--Lense effect, which causes a phase shift in the interferometer, in its full dynamical evolution.
\end{itemize}

A deep understanding of this topic could build,
on the way towards quantum gravity, a bridge
to the next level of interaction between quantum theory and gravity---the
level of quantum field theory in curved spacetime. Here, there exists
a definite prediction: black holes radiate with a thermal spectrum
(`Hawking effect'). In the special case of spherical symmetry
and absence of charges
(`Schwarzschild black hole'), this temperature is given by
\cite{Hawking}
\be
\lb{TSchwarz}
T_{\rm BH} =\frac{\hbar c^3}{8\pi k_{\rm B}GM}\approx 6.17\times 10^{-8}
 \left(\frac{M_{\odot}}{M}\right)\ {\rm K}\ ,
\ee
where $k_B$ denotes the Boltzmann constant and $M$ the mass of the black hole. This temperature is tiny
and has not yet been observed. It is only of
detectable size for sufficiently
small black holes, which can only form in the very early universe.
No such `primordial black holes' have, however, hitherto been seen.
There exists an effect in flat spacetime that is analogous to 
the Hawking effect. It arises for an observer who moves with constant
acceleration through Minkowski space. Such an observer experiences
the standard Minkowski vacuum as being filled with a thermal distribution
of particles, for which the temperature is given by \cite{Unruh}
\be 
\lb{TDU}
T_{\rm DU}= \frac{\hbar a}{2\pi k_{\rm B}c}\approx 4.05\times 10^{-23}
\ a\left[\frac{\rm cm}{{\rm s}^2}\right]\ {\rm K}\ ,
\ee
where $a$ denotes the acceleration. This temperature is often referred to as
the `Davies--Unruh temperature'. There are proposals 
(and attempts) to observe this
effect in particle accelerators or with ultra-intense lasers,
cf. \cite{Leinaas}. Of interest
are also suggestions to study {\em formal} analogues of these effects 
with features from condensed matter physics such as Bose--Einstein condensation,
cf. \cite{BLV} and the references therein.

For definiteness, we restrict ourselves in this study to GR and
standard quantum theory (plus unifications of both), and leave aside
alternative theories of gravity as well as, for example, non-linear
generalizations of quantum theory. Among such more general frameworks
are the suggestions that the gravitational field might be the cause
for a collapse of the wave function, cf. 
\cite{RP} and Chap.~8 in \cite{deco}.
This would signal a breakdown of unitarity at the most fundamental level.
There exist experimental proposals for a test of such ideas. For example,
the interaction of one photon with a tiny mirror could lead to
a superposition of mirror states, invoking in this process
about $10^{14}$ atoms \cite{mirror}.
The hope then will be that one may be able to distinguish between standard
environmental decoherence and a fundamental gravity-induced
collapse. At the moment, however, one is still very far away from 
an experimental test.

As for alternatives to classical GR, there are different options.
On the one hand, one may add additional fields in the gravitational 
sector, such as scalar fields. This leads to the Jordan--Brans--Dicke
theory and generalizations thereof \cite{FM}.
Typically, such theories lead to a spatiotemporal variation of
coupling constants, from which one gets strong constraints on them.
They play a role mainly as effective low-energy theories resulting
from fundamental theories such as string theory. On the other hand,
one may consider gauge theories of gravity. The simplest example is
the Einstein--Cartan theory, which invokes the concept of a non--vanishing
spacetime {\em torsion} (in addition to curvature), see, for example, 
\cite{Hehl} for a lucid introduction. In this theory, vierbein
(instead of metric) and connection arise as the gauge potentials
belonging to translations and Lorentz transformations, respectively.
The corresponding gauge field strengths are torsion and
curvature, respectively. The presence of torsion is needed in order
to accomodate spin (for scalars and photons, the Einstein--Cartan theory is
indistinguishable from GR). An even more general framework allows the 
presence of non--metricity, leading to metric--affine 
gauge theories of gravity,
 cf. \cite{NH}. As we shall mention below,
the effects of torsion turn out to be negligible on laboratory scales
(they may, however, play a role in the early universe),
so we can safely restrict ourselves to GR for the purpose of this study.


\section{Quantum mechanics in external electromagnetic and gravitational
         fields}

In order to study the interaction of quantum systems with gravity,
an appropriate formalism has to be developed. Since we are dealing here with
situations where quantum field-theoretic effects such as particle creation
can be neglected, it is sufficient to consider quantum mechanics
in an external gravitational field. In the non-relativistic limit,
the appropriate setting is the Schr\"odinger equation with an external
gravitational field. Since some experiments are also sensitive to
relativistic corrections, we discuss in the following 
the relativistic quantum-mechanical equations (Klein--Gordon and Dirac equation, respectively) and perform a non-relativistic
expansion. The Klein--Gordon equation describes spinless particles
(e.g. $\pi$--mesons), while the Dirac equation describes spin-1/2 particles.
For most applications, the latter is the appropriate framework.

\subsection{The Klein--Gordon equation and its non-relativistic expansion}

The Klein--Gordon equation in flat spacetime without external fields
is given by
\be
\lb{KG}
\left(\hbar^2\Box-m^2c^2\right)\varphi(x^{\mu})=0\ ,
\ee
where $m$ is the mass of the particle, and
\bdm
\Box\equiv-\frac{\partial^2}{c^2\partial t^2}+\Delta
\equiv\eta^{\mu\nu}\partial_{\mu}\partial_{\nu}\ .
\edm
The coupling to an external gravitational and electromagnetic field
is achieved through `minimal coupling', that is, through the substitution
of the partial derivative $\partial_{\mu}$ by the covariant derivative
$D_{\mu}$, which is defined by its action on a vector field $v^{\mu}$ as
\be
\lb{D}
D_{\mu}v^{\nu}=\partial_{\mu}v^{\nu}+\Gamma^{\nu}_{\mu\lambda}v^{\lambda}
-\frac{\I e}{\hbar c}A_{\mu}v^{\nu} \ ,
\ee
where $\Gamma^{\nu}_{\mu\lambda}$ denotes the Christoffel symbol, $A_{\mu}$ the electromagnetic vector potential, and $e$ denotes the charge of the particle.
Therefore, instead of (\ref{KG}) one has
\be
\lb{KGD}
\left(\hbar^2g^{\mu\nu}D_{\mu}D_{\nu}-m^2c^2\right)\varphi(x^{\mu})=0\ .
\ee
We assume the external fields to be sufficiently weak and slowly
varying in space and time, in order to have no quantum field-theoretic effects. The
Klein--Gordon equation (\ref{KGD}) is then interpreted as a quantum-mechanical
equation. For the purposes of this study it is appropriate to
perform a non-relativistic approximation, that is, an approximation
in powers of $c^{-2}$. This can technically be achieved by a
Foldy--Wouthuysen-type of transformation or by an expansion of the
phase of $\varphi$. Effectively, the formalism yields a separation into
`positive' and `negative' energies.
For the case of a pure electromagnetic field,
the approximation scheme can be found, for example, in \cite{BD}, and
for the presence of both external fields in \cite{CL2}.
The result is a Schr\"odinger equation
\be
\lb{schrodinger}
\I\hbar\frac{\partial}{\partial t}\psi(\X,t)=\tilde{H}_{\rm eff}\psi(\X,t)\ ,
\ee
where $\psi$ denotes a redefined positive part of $\varphi$, and
$\tilde{H}_{\rm eff}$ is the non-relativistic Hamilton operator
{\em plus} relativistic correction terms.

Since the external gravitational field is assumed to be weak,
a post-Newtonian approximation should be sufficient \cite{CL2}. 
A post-Newtonian
approximation is an expansion in powers of the small parameter
\be
\frac{\bar{v}}{c} \sim \left(\frac{G\bar{M}}{c^2\bar{r}}\right)^{1/2}\ ,
\ee
where $\bar{v}$, $\bar{r}$, and $\bar{M}$ are typical values for
the velocities, radii, and masses of the involved bodies, respectively,
and the virial theorem has been used. Gravitational radiation is
not included in this approximation.
One can then choose for the metric coefficients the expressions
\bea
\lb{post}
g_{00}&\approx& -\left(1-\frac{2U}{c^2}+\frac{2U^2}{c^4}\right)\ ,\nonumber\\
g_{0i}&\approx& 0\ ,\nonumber\\
g_{ij}&\approx& \left(1+\frac{2U}{c^2}\right)\delta_{ij}\ ,
\eea
where
\bdm
U(\X,t)=G\int\frac{\rho_{\rm m}(\X',t)}{\vert\X-\X'\vert}\D^3x'
\edm
is the Newtonian potential corresponding to a mass density
$\rho_{\rm m}(\X',t)$. Since the off-diagonal elements of the metric
are neglected, rotation effects cannot be seen in this approximation.
Since the Klein--Gordon equation (\ref{KGD}) leads to the conservation
of a scalar product different from that of the Schr\"odinger equation, the
standard inner product
\be
\lb{scalar}
\langle\psi\vert\chi\rangle=\int\D^3x\ \psi^*\chi
\ee
is not conserved with respect to the Hamiltonian $\tilde{H}_{\rm eff}$
in (\ref{schrodinger}). Conservation can only be obtained after a
{\em time-dependent} transformation of the Hamiltonian and the wave function.
After such a transformation has been performed, the effective 
Hamiltonian reads \cite{CL2}
\bea
\lb{Heff}
H_{\rm eff}& = & mc^2 -\frac{\hbar^2}{2m}\Delta-eA_0-mU-
\frac{\hbar^4}{8m^3c^2}\Delta^2\nonumber\\
& & +\frac{3\hbar^2}{2mc^2}(U\Delta+\nabla U\cdot \nabla)+\frac{mU^2}{2c^2}
+ \frac{3\hbar^2\Delta U}{4mc^2} \nonumber\\
& & -\frac{e}{4mc}\left[1+\frac{\hbar^2}{2m^2c^2}\Delta-
\frac{3U}{c^2},{\mathbf B}(\X_{\rm a},t) \, {\mathbf L}\right]_+\nonumber\\
& & -{\mathbf d}\left({\mathbf E}_{\rm T}(\X_{\rm a},t)+
\frac{1}{c}{\mathbf v}_{\rm a}\times{\mathbf B}(\X_{\rm a},t)\right)
+{\mathcal O}\left(\frac{1}{c^4}
\right) \ .
\eea
Here, $\X_{\rm a}$ is the expectation value of the position operator
(describing, for example, the position of an atom), ${\mathbf L}$
the angular momentum operator, ${\mathbf d}$ the electric dipole
operator, and ${\mathbf v}_{\rm a}=\D\X_{\rm a}/\D t$; 
${\mathbf E}_{\rm T}$ and ${\mathbf B}$ denote the transversal
part of the electric and magnetic field, respectively.
The Hamiltonian (\ref{Heff}) is Hermitian with respect to the scalar product
(\ref{scalar}). 

The first four terms represent the rest energy, the non-relativistic kinetic term, and
the electric and gravitational potentials, respectively, while the
fifth term is the first relativistic correction to the kinetic term.
The next four terms are relativistic corrections for the gravitational
field.
The term proportional to $\Delta U$ in (\ref{Heff}) can be interpreted
as a `gravitational Darwin term', in analogy to the usual electric
Darwin term $\propto \nabla\cdot{\mathbf E}$ arising from 
`zitterbewegung' (the coordinate of a particle is smeared out over
a length $\approx \hbar/mc$).
 The latter appears for the Klein--Gordon equation
first at order $c^{-4}$. The remaining terms contain the
usual terms for the electric and magnetic field
as well as a commutator term between the magnetic field and
the gravitational potential.
Thus, one recognizes that the interaction of the matter field
with the magnetic field is modified by the gravitational potential $U$,
whereas its coupling to the transversal electric field and to $A_0$
remains unchanged in this order of
approximation (compared to the gravity-free case). This is important
for atom interferometry.

\subsection{The Dirac equation and its 
non-relativistic expansion}\label{DiracEq}

The non-relativistic expansion 
of the Dirac equation in an external electromagnetic field
can be treated by a Foldy-Wouthuysen transformation \cite{BD}.
In this way one finds correction terms analogous to the
electromagnetic terms in (\ref{Heff})
but augmented by terms involving spin. They usually find applications
in atomic physics (modification of the spectral lines found from
the Schr\"odinger equation). Similarly, one can discuss the
Dirac equation for rotating and accelerating systems and, correspondingly, for
the case of an external gravitational field. For rotation
{\boldmath$\omega$} and acceleration ${\mathbf a}$,
the non-relativistic expansion has been performed in \cite{HN}, while in \cite{Obukhov}, an exact non-relativistic result has been obtained for static spacetimes.

The Dirac equation in Minkowski space
(and for Cartesian coordinates) reads
\be
\left({\rm i}\gamma^{\mu}\partial_{\mu}+
         \frac{mc}{\hbar}\right)\psi(x)=0\ ,
\lb{Dirac}
\ee
where $\psi(x)$ is a Dirac spinor, and
\be
[\gamma^{\mu},\gamma^{\nu}]_+\equiv\gamma^{\mu}\gamma^{\nu}
+\gamma^{\nu}\gamma^{\mu}=2\eta^{\mu\nu}\ .
\ee
The transformation into an accelerated frame is achieved by replacing
partial derivatives with covariant derivatives, see, for example,
\cite{honnef},
\be
\lb{1.15}
\partial_{\mu} \longrightarrow D_{\mu}\equiv
\partial_{\mu}+\frac{{\rm i}}{2}\sigma^{mk}\Gamma_{\mu mk}\ ,
\ee
where $\sigma^{mk}={\rm i}[\gamma^{m},\gamma^{k}]$
is the generator of the Lorentz group, and $\Gamma_{\mu mk}$ 
denotes the components of the connection. Latin indices
here denote anholonomic components, that is, components
that are not derivable from a coordinate basis. From the
equivalence principle, one would expect that the form
in an accelerated frame gives also the
appropriate form in curved space--time, where
\be
[\gamma^{\mu},\gamma^{\nu}]_+ =2g^{\mu\nu}\ .
\ee
For the formulation of the Dirac equation in curved space--time, one has
to use the tetrad 
(`vierbein') formalism, in which a basis
$e_{n}=\{e_0,e_1,e_2,e_3\}$ is chosen at each space--time point.
This is the reason why anholonomic components come into play.
One can expand the tetrads with respect to the tangent vectors
along coordinate lines (`holonomic basis') according to
\be
e_{n}=e^{\mu}_{n}\partial_{\mu}\ .
\ee
Usually one chooses the tetrad to be orthonormal,
\be
e_{n}\cdot e_{m}\equiv g_{\mu\nu}e^{\mu}_{n}
e^{\nu}_{m}=\eta_{nm}\equiv{\rm diag}(-1,1,1,1)\ .
\ee
The reason why one has to go beyond the pure metric formalism is the fact
that spinors (describing fermions) are objects whose wave components
transform with respect to a two-valued representation of the
Lorentz group. One therefore needs a local Lorentz group and local
orthonormal frames.  

One can define anholonomic Dirac matrices according to
\be
\gamma^{n}\equiv e_{\mu}^{n}\gamma^{\mu}\ ,
\ee
where $e^{\mu}_{n}e_{\mu}^{m}=\delta_{n}^{m}$.
This leads to
\be
[\gamma^{n},\gamma^{m}]_+ =2\eta^{nm}\ .
\ee
The Dirac equation in curved space--time or accelerated frames then reads
\be
\left({\rm i}\gamma^{n}D_{n}+\frac{mc}{\hbar}\right)\psi(x)=0\ .
\lb{Diraccurved}
\ee
In order to study quantum effects of fermions in the gravitational field
of the Earth, one specializes this equation to the non-inertial
frame of an accelerated and rotating observer, with acceleration
${\mathbf a}$ and angular velocity {\boldmath$\omega$}, respectively
(see e.g. \cite{honnef}). A non-relativistic approximation
with relativistic corrections is then obtained by the standard
Foldy--Wouthuysen transformation. This leads to
(writing $\beta\equiv\gamma^0$)
\be
{\rm i}\hbar\frac{\partial\psi}{\partial t}=H_{\rm FW}\psi\ ,
\ee
with
\bea
H_{\rm FW} &=& \beta mc^2 -\frac{\beta\hbar^2}{2m}\Delta 
+ \beta m({\mathbf a} {\mathbf x})
  -\frac{\beta\hbar^4}{8m^3c^2}\Delta^2 \nonumber\\
& & -\mbox{\boldmath$\omega$}({\mathbf L}+{\mathbf S})-
\frac{\beta\hbar^2}{2m}{\nabla}\frac{\mathbf{a\ x}}{c^2}{\nabla}
-\frac{\I\beta\hbar^2}{4mc^2}\mbox{\boldmath$\sigma$}({\mathbf a}\times
{\nabla})+ {\mathcal O}\left(\frac{1}{c^3}\right)
\lb{HFW}
\eea
({\boldmath$\sigma$} denotes the Pauli matrices). The interpretation of the
various terms in (\ref{HFW}) is straightforward. The first, second, and fourth
terms correspond to the rest energy, 
the usual non-relativistic kinetic term, and
the first relativistic correction to the kinetic term, respectively,
cf. the analogous terms in the effective Hamiltonian (\ref{Heff}) arising from the Klein--Gordon equation.
The third term gives an interaction with the acceleration that
is relevant for interference experiments in the 
gravitational field or for accelerated systems, see Sect.~2.3.
(It corresponds to $-mU$ in (\ref{Heff}).)
The term {\boldmath$\omega$}${\mathbf L}$ describes the Sagnac effect, while 
{\boldmath$\omega$}${\mathbf S}$ corresponds to a spin-rotation effect (`Mashhoon
effect') 
that cannot be found from the Schr\"odinger equation (but it can be found from
the Pauli equation, which arises from the Dirac equation in
the non-relativistic limit).
One can estimate that for typical values of a neutron interferometer
experiment, the Mashhoon effect contributes only $10^{-9}$ of the 
Sagnac effect. This is very small, but it has been indirectly
observed \cite{Mashhoon}.
The term after these rotation terms corresponds to the
term containing $\nabla U\cdot\nabla$ in (\ref{Heff}). (The term
$\propto U^2$ in (\ref{Heff}) has no analogue here because
it arose from the post--Newtonian order $\propto U^2$ in (\ref{post}), 
which is not considered here.)

The case of a constant gravitational acceleration ${\mathbf g}$ is obtained
from (\ref{HFW}) by replacing ${\mathbf a}$ with ${\mathbf g}$.
The effect of curvature has thus not yet been taken into account.
The Dirac equation in an external gravitational field in the
post-Newtonian approximation was discussed in \cite{fisch}. There, an
additional external electromagnetic field has been considered, although
not in a form that can be used directly for the discussion below.

If torsion were present in spacetime, it would give additional terms in
(\ref{HFW}). From an analysis of existing experiments dealing with
atomic spectra, one can get stringent limit on torsion; more precisely,
on the spatial component of the axial torsion, $K$, because only this
component enters in this limit. The result is 
$K\leq 1.5\times 10^{-15}\ {\rm m}^{-1}$ \cite{CLtorsion}. Torsion, even if
present on the fundamental level, can thus safely be neglected on
the scale of laboratory physics (provided, of course, that this
experimental limit holds for all components of the torsion). The Foldy-Wouthuysen transformation for an electron in an external electromagnetic field and in Minkowski space, but with non-vanishing torsion, is described in \cite{Shapiro}.

\subsection{Observed effects and theoretical ideas}

Observations of the interaction between quantum effects and gravity
have mainly been achieved---apart from
free-fall situations, for example, of freely falling Bose-Einstein condensates \cite{Ketterle}---in the realm of neutron and atom
interferometry. Colella, Overhauser, and Werner (`COW') have measured
in 1975 a phase shift in a neutron's wave function caused by
gravity \cite{COW}, while Bonse and Wroblewski have performed in 1983
the analogous experiment for neutrons in an accelerated frame \cite{BW}. The observed effect
can immediately be derived from the third term in (\ref{HFW}). 
The COW experiment has in particular confirmed the validity of
the weak equivalence principle for this quantum system.
The quantum Sagnac effect described by the term {\boldmath$\omega$}${\mathbf L}$
in (\ref{HFW}) has also been seen \cite{WSC}, as well as the Mashhoon effect. We are not aware of experiments exploiting the other correction terms in (\ref{HFW}).

Another application of neutron interferometry is the demonstration that
neutrons have a discrete energy spectrum in the gravitational field
of the Earth, as predicted by the Schr\"odinger equation \cite{grenoble}.
The minimum energy is $1.4\times 10^{-12}$ eV, which is much smaller
than the ground-state energy of the hydrogen atom. 
Other experiments use atom interferometry, since atoms are easier to
handle and allow higher precisions in the experiments, cf. \cite{CL3}.

We also mention that among the many applications of
future experiments in space one also attempts to study the influence of
gravity on quantum entanglement \cite{OB}. An investigation of the possible lack of an influence of gravity on the quantum Hall effect can be found in \cite{HOR}.

Finally, we note some further ideas which have led to proposed
candidates for a measurable interaction between quantum mechanics and 
GR: (i) the interaction of Rydberg atoms with 
gravitational fields \cite{PFP}, (ii) an 
atomic analogue of the LIGO interference experiment labeled the 
`Matter-wave Interferometric Gravitational-wave Observatory' (MIGO), 
in which quantum interference between atoms replaces the classical 
interference between light rays \cite{MIGO} (this proposition has been criticized recently in \cite{MIGOno}), and (iii) the identification of the dark energy in the universe with the quantum fluctuations as determined in noise measurements in Josephson junctions \cite{BM}.


\section{Gravitational waves}

\subsection{Classical discussion}\label{classWaves}

GR has---like electrodynamics---solutions describing
waves propagating with the speed of light (`gravitational waves'). 
While there exist exact solutions of Einstein's equations describing
such waves, it is sufficient for many purposes to describe gravitational
waves in the weak-field approximation, in which the waves propagate on
a given background spacetime.
 The reason is that one expects
from an observational point of view only weak gravitational waves.
Starting point is thus the following decomposition of the 
metric,\footnote{This formalism can be found in all standard texts
on GR, see, for example, \cite{MTW} or \cite{NS}.}
\be
\lb{expansion}
g_{\mu\nu}=\eta_{\mu\nu}+f_{\mu\nu}\ ,
\ee
in which one assumes that there exist coordinates such that the
components of $f_{\mu\nu}$ are small. Introducing the combination
\be
\bar{f}_{\mu\nu}\equiv f_{\mu\nu}-\frac{1}{2}
\ \eta_{\mu\nu}f^{\rho}_{\ \rho}\ ,
\ee
and imposing the gauge condition $\partial_{\nu}\bar{f}_{\mu}^{\ \nu}=0$,
one finds the linearized Einstein equations,
 \be
\lb{2.7}
\Box\bar{f}_{\mu\nu}=-\frac{16\pi G}{c^4}T_{\mu\nu}\ .
\ee
The linearized level can be used both for the description of solar-system
effects such as light deflection near the Sun and the description of
rapidly varying fields such as gravitational waves (the situation
discussed here). In the vacuum case ($T_{\mu\nu}=0$) one finds
plane-wave solutions describing waves that are transversal to the
direction of propagation and have only two independent degrees of
freedom (like electromagnetic waves). The two independent linear
polarization states are inclined towards each other by 45 degrees
and are usually referred to as the $+$ and the $\times$ polarization. The wave thus possesses helicity two, which in quantum theory leads to the spin-2 nature of the massless graviton.

Although energy in general relativity has no local significance,
one can (independent of any approximation) prove that a gravitational wave
carries away positive energy to regions far away from any radiating source
(the amount of energy that can be radiated away is limited by the
ADM energy, which is non-negative), see, for example, Sect.~11.2 in
\cite{Wald}. 
In the linear approximation studied here, one can find from
(\ref{2.7}) the physically relevant solution given by the retarded potential,
\be
\lb{retarded}
\bar{f}_{\mu\nu}(\X,t)=\frac{4G}{c^4}\int\D^3x'\ \frac{T_{\mu\nu}
(t-\frac{\vert\X-\X'\vert}{c},\X')}{\vert\X-\X'\vert}\ .
\ee
 From this solution one finds in the quadrupole approximation
the following `quadrupole formula', which was already derived by Einstein
in 1918, and which gives the emitted power $P$ as \cite{MTW,NS}
\be
\lb{quadrupole}
P\equiv-\frac{\D E}{\D t}=\frac{G}{5c^5}\sum_{i,j=1}^3
\frac{\D^3Q_{ij}}{\D t^3}\frac{\D^3Q_{ij}}{\D t^3}\ ,
\ee
where 
\be
Q_{ij}=\int\D^3x\ \left(x_ix_j-\frac{1}{3}\delta_{ij}r^2\right)
\rho_{\rm m} 
\ee
denotes the mass quadrupole tensor (evaluated
at retarded time). In the special case of a test mass $m$ that is freely
falling on a circular orbit 
in the Schwarzschild metric of a mass $M\gg m$, one finds from
(\ref{quadrupole}),
\be
\lb{schwarz}
P=\frac{c^5}{5G}\left(\frac{2GM}{rc^2}\right)^3
\left(\frac{2Gm}{rc^2}\right)^2\ .
\ee
The prefactor has the dimension of power, $c^5/5G\approx 7.29\times
10^{51}$ Watt, but this would only be released if $m$ could fall
close to the Schwarzschild radius corresponding to $M$. For the Earth moving around the
Sun, the power emitted in gravitational waves is only about
$200$ Watt. 

In laboratory situations the emission of gravitational waves is usually
estimated to be utterly negligible. Considering, for example, 
a rod of mass $M$ and length $L$ spinning about its centre with
frequency $\omega$, one finds from (\ref{quadrupole}) for the
emitted power,
\be
P=\frac{2G}{45c^5}M^2L^4\omega^6\ ,
\ee
which upon inserting typical values such as $M=1$ kg, $L=1$ m, 
$\omega=1\ {\rm Hz}$ yields the utterly small value
$P\approx 10^{-54}$ Watt. One would thus need a very strong gravitational
field (such as the one present in the coalescence of two compact objects)
to get a measurable effect. This has to be kept in mind when discussing
proposals aiming at generating sufficiently strong gravitational
waves in the laboratory through quantum effects (Sect.~3.2).

Gravitational waves have hitherto only been observed {\em indirectly}
through the decrease of the orbital period of the binary pulsar
PSR 1913+16 (and other pulsars as well).
 Attempts to {\em detect} (not generate) gravitational
waves on Earth are presently underway with the help of laser interferometers,
some of which have started to operate in 2002, cf. \cite{review}. 
These interferometers should be able to detect waves with an 
amplitude $h$, describing the relative change in separation of two test masses, of $h\sim 10^{-21}$ (for short signals) and down to
$h\sim 10^{-26}$ (for a signal of a duration of one year). They have
a frequency window from about 10 Hz to 10 kHz. The envisaged
space mission LISA (to be launched in 2012) will be able to measure
frequencies down to about $10^{-4}$ Hz. 

In addition to theses interferometers, there are also resonance cylinders
(`Weber bars') in use. The first detectors of this kind
have been developed already in the sixties. They have the disadvantage of being
sensitive only around a specific (resonant) frequency and to relatively
large amplitudes. It was estimated that the
efficiency of {\em emitting} gravitational radiation
 by such a cylinder is \cite{Weinberg}
\be
\eta_{\rm rad} = \frac{\Gamma_{\rm grav}}{\Gamma_{\rm heat}} 
= \frac{64 \, G M v_{\rm s}^4}{15 \, L^2 c^5 \, \Gamma_{\rm heat}} 
\approx 10^{-34}\ ,\label{weberApprox}
\ee
where $M$ is the mass of the bar, $L$ its length, $\Gamma_{\rm heat}$ is the decay rate due to dissipation,
and $v_{\rm s}$ is the speed of sound. 
In the classical (Weber bar) case, we have 
$v_{\rm s} \approx 10^{-5} c$. The emission of gravitational waves
by a Weber bar is thus utterly negligible. Suggestions about how
this may change by a coupling to quantum systems will be discussed
in the following subsection.

\subsection{Generation and detection of gravitational waves via quantum systems}

In the last section, the concepts of classical gravitational waves as solutions to the linearized Einstein equations were introduced. Despite the ongoing experiments to detect gravitational waves, the search has been unsuccessful so far, although the interferometers mentioned above are expected to lead to a detection in the future. In addition, there have been some approaches which suggest that the 
interaction of gravitational waves with quantum, instead of classical matter should be enhanced and would thus lead to better detection results. 
These ideas rest on certain properties of {\em quantum fluids} --- for example, superconductors, superfluids, quantum Hall (QH) fluids, and atomic Bose--Einstein condensates (BECs) --- exhibiting macroscopic coherence 
which is absent in classical materials due to the strong 
entanglement with the environment which leads to decoherence \cite{deco}.

The general interaction of gravitational fields with quantum fluids 
has been extensively studied. Already DeWitt \cite{DeWitt} and Papini 
\cite{Papini,Papini2,Papini3} investigated the interaction of 
Thirring--Lense fields with rotating superconductors and considered the 
induced quantum phase shift (see also Sect. \ref{supercons}). 
In \cite{Anandan}, Anandan and Chiao considered constructing 
antennas for gravitational radiation using superfluids, while 
\cite{Anandan2} considers the detection of gravitational radiation 
with superconducting circuits. In \cite{Peng}, Peng and Torr 
investigated the coupling of gravitational waves to superconductors, while Speliotopoulos found a cross-coupling term between the 
electromagnetic and gravitational field in the long-wavelength 
approximation, which could lead to a conversion between the 
two fields \cite{Spel}. Tajmar and de Matos considered a classical coupling 
between electromagnetic and gravitoelectromagnetic waves\footnote{Such a
coupling eventually has to be derived from the coupled Einstein--Maxwell
equations.} 
\cite{Tajmar,Tajmar2}, which, however, would lead to a very small 
interaction strength (cf. Sect. \ref{supercons}). 
Chiao {\em et al.} \cite{Chiao,Chiao4,Chiao3} considered using
quantum fluids as transducers, that is, 
as generator as well as detector of gravitational waves. 
Finally, the idea of conversion between gravitational and 
electromagnetic waves was recently discussed in \cite{Licht}.

We focus here on the ideas which describe the generation (and detection) of {\em gravitational waves} via the use of quantum fluids. First, the concept of a quantum fluid is discussed and its different properties in comparison to classical matter are addressed. Then, we investigate the arguments used in the research to suggest that quantum fluids should be better interaction partners of gravitational waves than classical materials. We consider the proposed coupling schemes and discuss their legitimacy. Finally, we discuss another investigation of the conversion between electromagnetic and gravitational waves as well as the experiments which have been conducted so far.

\subsubsection{Classical materials versus quantum fluids}\label{classQuant}

The important property of quantum fluids which is central to the interaction schemes proposed above is the existence of a macroscopic quantum-mechanical phase coherence which is not suppressed by interaction with the environment, such as, for example, phonons or photons in conductors and semiconductors. Macroscopic coherence is generated as a result of the condensation processes in the systems; this coherence is absent in classical materials.

In superconductors, the coherent ground state consists of a condensate in the form of Cooper pairs, separated from the excited states by an energy gap, the so-called BCS gap, see, for example, \cite{Abrikosov}. 
In the quantum Hall fluid, which is realized by a quasi-two dimensional electron gas confined by a strong magnetic field, the electrons form composite fermions with an even number of flux quanta to form a QH condensate at low temperatures. The ground state is energetically protected from decoherence by the QH gap \cite{Jain, Laughlin}. Superfluids and atomic BECs also show macroscopic phase coherence in the formation of condensates at low temperatures, see, for example,
 \cite{Pines}. As long as the transition to excited states is negligible, the coherences are not suppressed, and thus the reduced density matrix shows non-vanishing off-diagonal terms over spatially large distances (for this reason, one refers to these materials as quantum many-body systems with off-diagonal long-range order (ODLRO) \cite{Yang}).

It is the existence of these coherences which is argued to lead to a stronger interaction with gravitational fields and a more efficient generation (and detection) of electromagnetic and gravitational waves.

\subsubsection{General remarks about the (classical) interaction strength of gravitational waves with quantum fluids}\label{quant1}

In \cite{Chiao,Chiao4}, it is argued heuristically that gravitational waves should couple more strongly to quantum fluids than to classical materials because, due to the macroscopic phase coherence, the length scale of the material is (in principle) not limited to an acoustic wavelength $\lambda_{\rm s} = v_{\rm s} T$ 
of the material, where $v_{\rm s}$ is again the speed of sound 
and $T$ the period of the gravitational wave. Due to the classical coupling via acoustic waves, classical matter used as an antenna for gravitational waves is restricted to roughly the size $\lambda_{\rm s}$. In the quantum coupling, this restriction may not apply. Since the macroscopic quantum coherences in quantum fluids can in principle be of the same length scale as the gravitational wavelength $\lambda$, it is suggested that one could in principle, because the length restrictions do not apply in quantum fluids, replace the speed of 
sound $v_{\rm s}$ in the material by the speed of light $c$.

 From Sect. \ref{classWaves}, we know that for the efficiency of emitting gravitational radiation by a cylindrical bar, $\eta_{\rm rad} \sim v_{\rm s}^4$ (see (\ref{weberApprox})). Taking an approximate speed of sound in classical materials of $v_{\rm s} \approx 10^{-5} c$, Chiao's argument suggests an increased radiation emission efficiency of $({c}/{v_{\rm s}})^4 \approx 10^{20}$. It should be noted here that, in contrast to the classical considerations of gravitational-wave detection, where (astrophysical) frequencies in the range of Hz to kHz are considered, the frequencies, corresponding to the above arguments of wavelengths on the same length scale as the detector, would be of the order of 10 MHz and higher (see Sect. \ref{experiment} for an experiment conducted to this end). No detectable astrophysical sources are available in this 
frequency range.

Furthermore, it is suggested that through `impedance' matching of the gravitational wave to the quantum fluid, an enhanced conversion between the two fields can be obtained. In the electromagnetic case, the impedance of free space is given by 
\be
Z_0 = \sqrt{\frac{\mu_0}{\eps_0}} \approx 377~\Omega~,
\ee
where $\eps_0$ and $\mu_0$ are the electric permittivity and magnetic permeability, respectively.
Analogously, the `impedance' for the gravitational wave is given by\footnote{It is shown in Sect. \ref{gravito} that the linearized Einstein equations can be written in a form analogous to Maxwell's equations, thus allowing for the definition of gravitational analogues of the electric permittivity and magnetic permeability. In the definition of the `impedance' for the gravitational wave, we have used SI units in order to show the analogy to the electromagnetic case. Note that in comparison to \cite{Chiao4}, the gravitational `impedance' $Z_{\rm g}$ defined above is smaller by a factor 4, which is due to the different definitions of the gravitoelectromagnetic potentials, see Sect. \ref{gravito}. }
\be
Z_{\rm g} = \sqrt{\frac{\mu_{\rm g}}{\eps_{\rm g}}} = \frac{4 \pi G}{c} \approx 2.79 \times 10^{-18}~{\rm m^2/kg~s}\ ,
\ee
where ${\eps_{\rm g}}$ and ${\mu_{\rm g}}$ are the gravitational permittivity and permeability of free space, respectively. Considering an electromagnetic wave impinging normally on a thin resistive film, the absorption, transmission, and reflection coefficients depend on the relationship between the resistance of the material and the impedance of free space $Z_0$ \cite{Born}. It is now argued that, solving a similar boundary problem for the gravitational wave, an analogous `impedance' matching condition can be derived. While in classical materials, the `resistance' is generally much higher than $Z_{\rm g}$, leading to virtual transparency with respect to the gravitational wave, quantum materials such as superconductors could have a `resistance' comparable to that of $Z_{\rm g}$ which could lead to maximal absorption of the gravitational wave. If the absorption could be viewed as an effective dissipation where gravitational-wave energy is converted into electromagnetic-wave energy instead of into heat, this would imply a reduced dissipation $\Gamma_{\rm heat}$, leading to an enhanced conversion process \cite{Chiao4}. These expectations are, however,
based on heuristic considerations and not on detailed calculations.\\

From the above arguments, one could expect that couplings between gravitational fields and quantum fluids be enhanced compared to the coupling to classical matter. However, even with the increase of the efficiency, as argued by Chiao, it is questionable whether this leads to effects which are large enough to be detected. Furthermore, it is important to note that the arguments are based on the classical efficiency formula derived by Weinberg \cite{Weinberg}. Since this derivation is based on classical properties of the detector such as sound waves in the material, it is not clear whether these arguments can be applied to quantum fluids. Generally, a theory regarding the quantum nature of the (macroscopic) coherences would have to be considered. It should be noted that (classical) Weber bars have been studied in the quantum regime by using quantum nondemolition measurements (see e.g. \cite{Braginsky,Braginsky2}). A quantum measurement analysis is
also crucial for the detection of gravitational-wave signals in current
interferometer experiments \cite{BChen}.

\subsubsection{Quantum-mechanical coupling considerations}\label{quant2}

Now we want to discuss the quantum-mechanical arguments used to explain the coupling between the electromagnetic and gravitational fields. Again, these are based on the existence of macroscopic phase coherence that leads to the generation of electromagnetic radiation and, by means of time reversal, to the detection of electromagnetic waves by generating gravitational waves \cite{Chiao}.

It is well known that, in the presence of an electromagnetic potential $A_\mu$, the quantum-mechanical wave function of a charged particle picks up a phase (the Aharonov--Bohm phase \cite{AB}),
\be
\Psi({\bf x}) \rightarrow \Psi({\bf x})~\exp \left(\frac{\I e}{\hbar c} \int \!\! A_\mu \D x^\mu \right)\ ,
\ee
which has been experimentally detected by means of interference experiments, see, for example, \cite{Chambers}. More generally, if a system evolves adiabatically (i.e., if the quantum adiabatic theorem holds) in a parameter space $\{\bf R\}$ around a closed curve $C$ in the presence of a non-trivial topology, the wave function accumulates a (topological) Berry phase \cite{Berry,Shapere},
\be
\Psi_n({\bf x}) \rightarrow e^{\I \gamma_n(C)}\Psi_n({\bf x})\ ,
\ee
where the phase is explicitly given by
\be
\gamma_n(C) = \I \oint\limits_C \scProdtoo{n({\mathbf R})}{\nabla_{\mathbf R} n({\mathbf R})} \cdot \D{\mathbf R}\ ,
\ee
where $\ket{n({\bf R})}$ are eigenstates of the system.
The Berry phase has been experimentally detected, for example, by considering the polarization of neutrons in the presence of a spiral magnetic field \cite{Bitter} (see also \cite{Tomita,Delacretaz}). 

In the Aharonov--Bohm case, a time-dependent phase leads to an `electromotive' force which acts on the electrons. Similarly, a Berry phase can, in the case of a coupling of the spin with an external field, lead to a `(nonelectro)motive' force which acts on the spin of the particle. In both cases, this is a result of the fact that the free energy of the system is dependent on the quantum-mechanical phase; the result is an electric, non-dissipative equilibrium (persistent) current in the mesoscopic system \cite{Stern} (for an experimental measurement of the Aharonov--Bohm persistent current, see e.g. \cite{Webb}).

It has been shown by DeWitt \cite{DeWitt} and Papini \cite{Papini2} that in the case of weak gravitational fields and slow velocities (not suitable to describe gravitational waves), the so-called gravitomagnetic potential ${\bf A}_{\rm g}$ (cf. Sect. \ref{gravito}), arising from the gravitational field, leads to a Berry-type phase
\be
\Delta S \sim \oint {\bf A}_{\rm g} \cdot \D{\bf l}\ .
\ee
It is now argued by Chiao {\em et al.} \cite{Chiao} that in the presence of a {\em gravitational wave}, the electron wave function should also pick up a Berry-type phase which, due to its spin, would lead to a macroscopic electric current which, in turn, would produce electromagnetic waves. Assuming time-reversability, the opposite process of the conversion of electromagnetic waves into gravitational waves should also be possible.\\

Now let us turn to the mathematical model involved in the coupling. The different models which are discussed in the literature consider the case of minimally coupled electromagnetic and gravitational fields. The classical derivations, considered, for example, by DeWitt \cite{DeWitt} and elaborated by Papini \cite{Papini,Papini2}, start from the Lagrangian of a `classical' electron in an electromagnetic and weak gravitational field (see Sect. \ref{before} for details),
without considering gravitational waves. This leads to a Hamiltonian with cross-coupling terms of the form
\be
H_{\rm int} \sim {\bf A} \cdot {\bf A}_{\rm g}\ ,
\ee
where ${\bf A}$ and ${\bf A}_{\rm g}$ are the magnetic and gravitomagnetic 3-potentials. Speliotopoulos also obtained terms of the above form in his long-wavelength approximation model \cite{Spel}.

Coupling of the electron including its quantum-mechanical properties via minimal coupling in the Dirac equation as in Sect. \ref{DiracEq}, but including here the electromagnetic field as well, is achieved by (cf. Sect. \ref{DiracEq} for details on the index notation)
\be
p_\mu \rightarrow p_\mu - \frac{e}{c} A_\mu + \frac{\hbar}{2} \sigma^{m k} \Gamma_{\mu m k}\ .\label{relat}
\ee
For the purely gravitational case, the non-relativistic expansion of this Dirac equation coincides with the prescription of
\cite{HN}. As far as we know, there has not been a general systematic derivation of the non-relativistic Hamiltonian in the presence of both electromagnetic and gravitational fields, neither in a low-order post-Newtonian approximation such as is used in the derivation of (\ref{HFW}), nor in the background of a gravitational wave. The results of \cite{fisch} employ a post-Newtonian approximation (where gravitational waves are not seen) with an application to the hydrogen
atom. It is thus questionable whether the non-relativistic result will contain, as proposed by Chiao {\em et al.} \cite{Chiao}, in the effective Hamiltonian a term of the form
\be
H = \frac{1}{2 m} (p_i - \frac{e}{c} A_i + G_i)^2 + V\ ,
\ee
where $G_i$ denotes the non-relativistic form emerging from the minimal coupling to the gravitational field in (\ref{relat}). On algebraic grounds, it could certainly be possible that the resulting coupling terms between the electromagnetic and gravitational fields vanish due to the properties of the $\gamma$-matrices. It is, however, highly recommended that this derivation be performed in order to gain insight into the possibility of such a coupling as well as its strength.

If one assumes the above form for the interaction Hamiltonian, one is led directly to cross-coupling terms
\be
H_{\rm int} \sim {\bf A} \cdot {\bf G}\ ,\lb{47}
\ee
where ${\bf A}$ is the magnetic 3-potential and ${\bf G}$ incorporates the gravitational field; in comparison to the results derived by DeWitt and Papini, these terms would include the quantum-mechanical coupling between the gravitational and electromagnetic field via the spin of the electron.
It should be noted here that in a conversion process between electromagnetic and gravitational fields described by (\ref{47}), momentum conservation must be obeyed, either by the addition of momentum as supplied by the medium (e.g. strong magnetic fields in the quantum Hall fluid) or by considering electromagnetic quadrupole fields (e.g. in the conversion processes involving superconductors) \cite{Chiao3}.

\subsubsection{Experimental and quantitative theoretical results}\label{experiment}

We know of only one experiment which has been carried out in a laboratory framework where a quantum fluid is supposed to act as a transducer between gravitational and electromagnetic waves. In this experiment \cite{Chiao3}, superconducting yttrium barium copper oxide (YBCO) was used as a transducer. Two probes were used, one as a generator and the other as a detector of gravitational waves. Electromagnetic quadrupole radiation with a frequency of 12 GHz is shone upon the first transducer, producing gravitational radiation which falls upon the second superconductor, where the generated electromagnetic quadrupole radiation is detected. In order to prevent direct electromagnetic coupling, the two superconductors are enclosed by Faraday cages.

At least in this relatively simple setup, there has been no detection of generated gravitational waves, which, according to Chiao {\em et al.} \cite{Chiao3}, gives an upper conversion efficiency of $\eta < 1.6 \times 10^{-5}$ (if one assumes equal conversion efficiencies from one form of wave to another). Chiao {\em et al.} argue that this negative result may be due to
\bi
\item possible high residual microwave and far-infrared losses in YBCO, or
\item generally large ohmic dissipation arising from high temperatures; this could be reduced at lower temperatures.
\ei
It is possible that the experimental conditions could be improved for this situation. However, since there is as yet no evidence for a coupling of such strength as to be detectable, it would be preferable first to investigate that matter theoretically.\\

Finally, we want to shortly comment on calculations performed by Licht \cite{Licht}. He considers gravitational waves impinging on normal conductors as well as superconductors and calculates, using classical electrodynamical arguments, the generation efficiency of outwardly propagating electromagnetic waves.  The superconductor is described within a phenomenological Ginzburg-Landau approach (see Sect. \ref{GL}). The result is that the generation of electromagnetic waves in superconductors in both the low and high frequency domain is not strongly enhanced compared to that in normal conductors and should not lead to detectable results. It should be noted, however, that in comparison to the argumentation of Chiao {\em et al.}, no quantum-mechanical properties such as the spin of the electron have been considered in these calculations.

\subsubsection{Conclusion on the attempts to generate and detect electromagnetic waves in quantum systems}\label{concl1}

To summarize the proposed ideas of using quantum fluids as transducers between electromagnetic and gravitational waves:\\

There have been arguments presented for the case of a stronger coupling of gravitational waves with quantum fluids than is the case with classical materials. Heuristic considerations of Weinberg's formula (\ref{weberApprox})
for the emission efficiency of gravitational waves from a (classical) Weber bar have been presented to give enhancements for quantum fluids. Yet, it is not clear as to whether these arguments can be applied to this formula since its derivation is based on classical properties. As far as the arguments based on the Dirac equation are concerned, to our knowledge there is no non-relativistic derivation available which includes both electromagnetic and gravitational field in a general post-Newtonian approximation (which in low order does not describe gravitational waves) or in the background of a gravitational wave. In order to make qualitative as well as quantitative comments on the coupling strength, it is highly recommended to perform this derivation.

The theoretical investigations performed by Licht \cite{Licht} lead to generation efficiencies which are so small as not to be relevant for experimental laboratory research. Since the experiments conducted by Chiao {\em et al.} have also not yet led to positive results, it would be advisable that the interaction first be further explored on a theoretical basis before new experiments are carried out.


\section{Gravitational effects in superconductors}\label{supercons}

In the last section, we have discussed proposed interaction mechanisms and experiments between gravitational and electromagnetic fields, where the emphasis was on the use of quantum systems as detector and generator of {\it gravitational waves}. Here, we will discuss interaction processes involving weak gravitational fields coupled to rotating superconductors, where the emphasis is on small velocities, thus specifically {\em excluding} gravitational waves. We will review the work which has been done in this area so far and the different mechanisms which have been proposed and give a critical discussion of the ideas and the possible realizations and measurements of such interactions.

\subsection{Foundations}

To begin, it is necessary to present the basic concepts which are used. First, we introduce the concept of {\em gravitoelectromagnetism}, which follows in the weak-field approximation of GR for slowly moving matter. This is a valid approximation as long as we are concerned with laboratory experiments which do not deal with gravitational waves. We will then mention a proposed classical coupling scheme between electro- and gravitoelectromagnetism which has been discussed in the literature. Finally, we introduce the concepts for superconductors which are important for the argumentations used in the current approaches.

\subsubsection{Gravitoelectromagnetism}\label{gravito}

Within the linear approximation of GR, gravitational analogues of the electric and magnetic fields of Maxwell's theory of electromagnetism can be derived if we assume, in addition, that we are only considering slowly moving objects \cite{LT,Forward,Mashhoon2}. The field equations can be formulated in a Maxwell-type structure. In this section, we will mention the most important concepts of this so-called theory of gravitoelectromagnetism, using the notation of \cite{Mashhoon2}.\footnote{There is a certain degree of arbitrariness in the definition of the gravitoelectromagnetic potentials. The notation used here is chosen as to clearly show the analogy to the electromagnetic case, while also pointing out the difference (as witnessed in the factor 1/2 for the gravitomagnetic field which stems from the ratio of gravitomagnetic to gravitoelectric charge $q_{\rm B}/q_{\rm E} = 2$).}

The following derivation is similar to that for gravitational waves (see Sect. \ref{classWaves}). It is assumed that only weak perturbations of flat spacetime occur, so that we may again consider $g_{\mu \nu} = \eta_{\mu \nu} + f_{\mu \nu}$, where $f_{\mu \nu}$ is small. The Einstein equations then reduce to (\ref{2.7}). However, in contrast to the derivation of gravitational waves, we take here as a source an ideal fluid, assuming velocities small compared to $c$: $T^{0 0} = \rho_{\rm m} c^2$, $T^{0 i} = c j^i$, and $T^{i j} = {\mathcal O}(c^0)$, where $\rho_{\rm m}$ is the matter density and ${\bf j} = \rho_{\rm m} {\bf v}$ is the matter current density.

If we now introduce the (scalar) gravitoelectric and (vector) gravitomagnetic potentials,
\be
\Phi_{\rm g} := \frac{c^2}{4} \bar{f}_{0 0}\ , \qquad (A_{\rm g})_i := -\frac{c^2}{2} \bar{f}_{0 i}\ ,\label{defPot}
\ee
respectively, we can write the Einstein equations up to ${\mathcal O(c^{-4})}$ in the following form:\\\\
\hspace*{2cm}
\begin{tabular}[t]{l l}
$\nabla \cdot {\bf E}_{\rm g} = 4 \pi G \rho_{\rm m}$  &$\nabla \times ({\frac{1}{2} \bf B}_{\rm g}) = \frac{1}{c} \partial_t {\bf E}_{\rm g} + \frac{4 \pi G}{c} {\bf j}$\\
$\nabla \times {\bf E}_{\rm g} = -\frac{1}{c} \partial_t ({\frac{1}{2} \bf B}_{\rm g})$  &$\nabla \cdot ({\frac{1}{2} \bf B}_{\rm g}) = 0$\ .\\\\
\end{tabular}\\
Here, we have introduced the gravitoelectric and gravitomagnetic fields
\begin{eqnarray}
&&{\bf E}_{\rm g} = -\nabla \Phi_{\rm g} - \frac{1}{c} \partial_t (\frac{1}{2} {\bf A}_{\rm g})\ ,\\
&&{\bf B}_{\rm g} = \nabla \times {\bf A}_{\rm g}\ ,
\end{eqnarray}
respectively.

It should be noted that the gravitoelectric and gravitomagnetic potentials defined above are just the Newtonian potential and `Thirring--Lense potential', respectively. This can be seen by regarding the solutions to (\ref{2.7}), given by (\ref{retarded}),
\be
\bar{f}_{\mu\nu}(\X,t)=\frac{4G}{c^4}\int\D^3x'\ \frac{T_{\mu\nu}
(t-\frac{\vert\X-\X'\vert}{c},\X')}{\vert\X-\X'\vert}\ .
\ee
If one considers a source distribution confined around the spatial coordinate origin, the potentials far away from the source can be written as
\be
\Phi_{\rm g} \sim \frac{G M}{r}\ , \qquad {\bf A}_{\rm g} \sim \frac{G}{c} \frac{{\bf J} \times \X}{r^3}\ ,
\ee
where $r = |\X|$, and $M$ and ${\bf J}$ are the mass and angular momentum of the source, respectively.\\

In order to point out the difference between linearized GR and electromagnetism, we remark that within the above approximations, it is possible to derive a Lorentz-type force in the lowest order of ${\bf v}/c$ and assuming ${\bf A}_{\rm g}$ is time-independent, yielding
\be
{\bf F} = -m {\bf E}_{\rm g} - 2 m \frac{{\bf v}}{c} \times {\bf B}_{\rm g}\ .
\ee
Introducing, in analogy to electromagnetism, the gravitoelectric and gravito\-magnetic charges $q_{\rm E} = -m$ and $q_{\rm B} = -2 m$, respectively, we have the ratio $q_{\rm B}/q_{\rm E} = 2$, compared to $q_{\rm B}/q_{\rm E} = 1$ in the electromagnetic case. This can be traced back to the fact that linearized gravity is a spin-2 theory, whereas electromagnetism is fundamentally spin-1. For a rotating body of mass $M$ that acts as a source, the gravitational charges are given by $q_{\rm E} = M$ and $q_{\rm B} = 2 M$ to preserve the attractive nature of GR. \footnote{Here, the Gaussian system of units has been used; however, it is helpful to bring the gravitoelectromagnetic field equations into SI form in order to introduce the concepts of the gravitoelectric permittivity and gravitomagnetic permeability. In order to define the gravitoelectromagnetic fields in an analogous way as in electromagnetism,
\bc${\bf E}_{\rm g} = -\nabla \Phi_{\rm g} - \partial_t {\bf A}_{\rm g}\ , \qquad {\bf B}_{\rm g} = \nabla \times {\bf A}_{\rm g}$\ ,\ec
it is necessary to define the potentials in a different way (see \cite{Tajmar}),
\bc$\Phi_{\rm g} := \frac{c^2}{4} \bar{f}_{0 0}\ , \qquad (A_{\rm g})_i := -\frac{c}{4} \bar{f}_{0 i}$\ .\ec
The SI form of the field equations then reads
\bc\begin{tabular}[t]{l l}
$\nabla \cdot {\bf E}_{\rm g} = \frac{\rho_{\rm m}}{\eps_{\rm g}}$  &$\nabla \times {\bf B}_{\rm g} = \frac{1}{c^2} \partial_t {\bf E}_{\rm g} + \mu_{\rm g} {\bf j}$\\
$\nabla \times {\bf E}_{\rm g} = -\partial_t {\bf B}_{\rm g}$  &$\nabla \cdot {\bf B}_{\rm g} = 0$\ ,
\end{tabular}\ec
with the gravitoelectric permittivity and gravitomagnetic permeability
\bc$\eps_{\rm g} = \frac{1}{4 \pi G}$ and $\mu_{\rm g} = \frac{4 \pi G}{c^2}$\ ,\ec
respectively.}

\subsubsection{Classical coupling between gravitoelectromagnetism and electromagnetism}\label{classical}

Based on the above analogy, a coupling between the electromagnetic and gravitoelectromagnetic fields can be derived \cite{Tajmar,Bertolami}. The matter density $\rho_{\rm m}$ and charge density $\rho$ of a point-like or uniformly distributed mass and charge configuration of mass $m$ and charge $e$ obey
\begin{equation}
\rho_{\rm m} = \frac{m}{e} \rho\ .
\end{equation}
From this, the following relationships between the potentials can be derived,
\begin{eqnarray}
&&\Phi_{\rm g} = G \frac{m}{e} \Phi \equiv \kappa_{\rm E} \Phi\ ,\label{coupling1}\\
&&{\bf A}_{\rm g} = G \frac{2m}{e} {\bf A} \equiv \kappa_{\rm B} {\bf A}\ .\label{coupling2}
\end{eqnarray}
Thus, a coupling can be defined as\footnote{It should be noted that in \cite{Bertolami}, the coupling constants have been defined in SI units and are given by $\kappa_{\rm E(B)} = 4 \pi G \eps_0 \frac{q_{\rm E(B)}}{e} \approx 7.41 \times 10^{-21} \frac{q_{\rm E(B)}}{e} ~~ {\rm C^2/kg^2}$\ .}
\begin{equation}
\kappa_{\rm E(B)} = G \frac{q_{\rm E(B)}}{e} \approx 6.67 \times 10^{-8} \frac{q_{\rm E(B)}}{e} ~~ {\rm cm^3/g~s^2}\ ,
\end{equation}
where $q_{E(B)}$ are the gravitoelectric and gravitomagnetic charges of the source, respectively. This leads to relationships between the electromagnetic and gravitoelectromagnetic fields,
\begin{eqnarray}
&&{\bf E}_{\rm g} = \kappa_{\rm E} {\bf E}\ ,\label{coupling3}\\
&&{\bf B}_{\rm g} = \kappa_{\rm B} {\bf B}\ .\label{coupling4}
\end{eqnarray}
Due to this coupling, the motion of a particle with both mass and charge should at the same time produce an electromagnetic as well as a gravitoelectromagnetic field. Also, if one combines eqs. (\ref{coupling3}) and (\ref{coupling4}) with Maxwell's equations, gravitoelectromagnetic fields should generate electromagnetic fields and vice versa.

Note, however, that even for increased values of the ratio $m/e$, for example, for moving ions, this coupling is tiny, since $\kappa_{\rm E(B)} \ll m/e$.

\subsubsection{Theories of superconductivity}\lb{GL}

This section will be devoted to a short overview of the important concepts involving superconductors insofar as they will be applied in the upcoming discussions. For further information on these theories, we refer to \cite{Abrikosov, London, Cabrera}. Since the argumentations in the current experiments are based on the London, and more generally, the Ginzburg--Landau, theory, we will restrict ourselves mainly to phenomenological descriptions of superconductivity, making only a few comments on the microscopic BCS theory. Furthermore, we will restrict ourselves to type I superconductors, where the Meissner effect is strictly valid up to the critical magnetic field where superconductivity is destroyed.\\

London set up a phenomenological theory which explains the two unique features of a superconductor, the Meissner effect and the loss of resistance below a critical temperature $T_c$. The input consists of Maxwell's equations 
together with the two `material equations'
\bea
\partial_t (\Lambda {\bf j}_{\rm s}) = {\bf E}\ ,\\
\nabla \times (\Lambda {\bf j}_{\rm s}) + \frac{1}{c} {\bf B} = 0\ ,
\eea
relating the current density ${\bf j}_s$ of the superconducting electrons and the electromagnetic fields in the superconductor. Here, $\Lambda=m/e^2n_{\rm s}$, where $n_{\rm s}$, $m$, and $e$ are the density, the mass, and the charge of the superconducting electrons, respectively, is related to the London penetration depth $\lambda$---determining the degree to which
a magnetic field can penetrate into a superconductor---by $\lambda = ({\Lambda c^2}/{4 \pi})^{1/2}$ \cite{Abrikosov}.

If one takes a multiply connected superconducting material and considers the flux through an area $S$ bordered by a curve $C$ around a hole, one finds 
(from the single-valuedness of the wave function) that
\be
\lb{quantcond}
\oint\limits_S {\bf B} \cdot \D{\bf S} + c \oint\limits_C \Lambda {\bf j}_{\rm s} \cdot \D{\bf l} = n \frac{h c}{2 e}\ ,
\ee
where $n$ is an integer which depends on the initial conditions of the superconducting system. This result can also be derived by applying the Bohr--Sommerfeld quantization condition to the local mean value of the canonical momentum ${\bf p}_{\rm s} = m {\bf v}_{\rm s} + \frac{e}{c} {\bf A}$ of the superconducting electrons around a closed path,\footnote{The factor 2 on the 
right-hand side of (\ref{quantCond}) arises from the fact that the superconducting electrons occur in pairs (Cooper pairs) and thus carry charge $2 e$ and mass $2 m$.}
\be
\oint\limits_C (m {\bf v}_{\rm s} + \frac{e}{c} {\bf A}) \cdot \D{\bf l} = n \frac{h}{2}\ ,\label{quantCond}
\ee
where ${\bf v}_{\rm s}$ is the velocity of the superconducting electrons, related to the current density by ${\bf j}_{\rm s} = n_{\rm s} e {\bf v}_{\rm s}$.

If the above curve runs several penetration depths in the material, the current density ${\bf j}_{\rm s}$ due to the superconducting electrons vanishes, and one obtains magnetic flux quantization through the area $S$. If the body is simply connected, the flux vanishes (Meissner effect).

If, on the other hand, one takes a superconducting body and {\em rotates} it, a (homogeneous) magnetic field is induced along the axis of rotation. This is caused by a weak surface current as a result of the superconducting electrons lagging behind the positive rest charges of non-superconducting electrons and nuclei (see \cite{Rystephanick} for an explanation of this effect in terms of Coriolis forces). The magnetic field induced in a body rotating with angular velocity {\boldmath$\omega$} can be calculated from the quantization condition (\ref{quantcond}) to be \cite{London,Verheijen}
\be
{\bf B} = -\frac{2 m c}{e} \mbox{\boldmath$\omega$}\ \label{LonMoment}
\ee
deep inside the superconducting material; it is called the {\em London moment}. In this derivation, it is used that for the initial condition of vanishing external magnetic fields and the superconducting body at rest, $n = 0$ occurs in the quantization condition.\\

More generally, the theory of superconductivity can be considered within the Ginzburg-Landau theory, which reproduces the London equations \cite{Abrikosov}. Integrating the total current density ${\bf j}$ around a closed loop, and including the effects of rotation, one obtains \cite{Cabrera,Capellmann}
\be
\oint\limits_{S} {\bf B} \cdot d{\bf S} + \frac{m c}{e^2 n_{\rm s}} \oint\limits_C {\bf j} \cdot \D{\bf l} + \frac{2 m c}{e} \mbox{\boldmath$\omega$} \cdot {\bf S} = n \frac{h c}{2 e} \ ,
\lb{additional}
\ee
where $C$ is a curve within the superconductor with associated surface $S$, and $n_{\rm s}$ is again the density of the superconducting electrons. 
The third term has arisen in addition to (\ref{quantcond}).
It can easily be seen that for an area $S$ bordered by a curve which lies many penetration depths in the superconductor, where the current density ${\bf j}$ vanishes (this can be fulfilled e.g. for a {\em thick} superconducting ring), and in the case of an initially non-rotating superconductor ($n$=0), the above equation again leads to the London moment. For a {\em thin} superconducting ring, it is found that for every $n$ there exists an angular velocity $\omega_n$ where the current density ${\bf j}$ and the magnetic flux $\oint\limits_S {\bf B} \cdot \D{\bf S}$ vanish together \cite{Cabrera}, leading to
\be
\frac{h}{2 m} = 2 S \Delta \omega\ ,\label{massCP}
\ee
where $\Delta \omega = \omega_{n+1}-\omega_n$; this equation may be used to determine the mass of the Cooper pairs $m' = 2 m$ \cite{Tate} (see also \cite{Cabrera} for relativistic mass corrections).\\

On a microscopic level, superconductivity can be explained by the theory first devised by Bardeen, Cooper, and Schrieffer (`BCS' theory) \cite{BCS} (for an introduction, see \cite{Abrikosov}). Within this theory, the ground state of a superconductor is described as a condensate of Cooper pairs, pairs of electrons which are bound by an {\em effective} electron-electron interaction mediated by the electron-phonon interaction of the electrons with the lattice. The theory also derives the so-called BCS gap between the ground and excited states of the superconductor, which leads to macroscopic coherence in the system (see Sect. \ref{classQuant}). The Ginzburg-Landau theory can be derived from the BCS theory, if (i) the temperature of the system is close to the critical temperature $T_c$ and (ii) the London penetration depth $\lambda$ is much greater than the coherence length $\xi$ of the Cooper pairs.

\subsection{Rotating Superconductors}

Now we will discuss the effects of gravitational fields on rotating superconductors in order to review the proposed effects and to assess the planned experiments and measurement possibilities.

\subsubsection{Earlier considerations}\label{before}

As mentioned above, London \cite{London} showed that if a superconducting sphere is set in rotation, a magnetic field is generated along the axis of rotation, which is a result of the momentum quantization condition. In this derivation, only the electromagnetic properties of the electrons have been used. The effect has been experimentally confirmed, see, for example, \cite{Hildebrandt,Sanzari,Verheijen}.\\

DeWitt, in a slightly different setup, addressed the {\em gravitational drag effect} on a superconductor \cite{DeWitt}. He considered a (thick) superconducting ring at rest and an axially symmetric mass placed in its centre. This mass is set in rotation, generating a Thirring--Lense field which acts on the superconductor (see \cite{Ross} for a similar result in covariant form).

He starts from the classical Lagrangian of an electron in electromagnetic and gravitational fields,
\be
L = -m c (-g_{\mu \nu} \dot{x}^\mu \dot{x}^\nu)^{1/2} + \frac{e}{c} A_\mu \dot{x}^\mu\ .
\ee
This is a purely classical model and does not capture quantum-mechanical spin effects (cf. Sect. \ref{quant2}). Using the approximation of weak gravitational fields and small velocities (cf. Sect. \ref{gravito}), the Hamiltonian for the many-electron system reads
\be
H = \sum\limits_n \{ (\frac{1}{2m} [{\bf p}_n - {\bf F}(x_n)]^2 + V(x_n) \} + V_{\rm int}\ .\lb{68}
\ee
Here, the generalized magnetic vector potential ${\bf F} = \frac{e}{c} {\bf A} - \frac{2 m}{c} {\bf A}_{\rm g}$ includes the magnetic as well as the gravitomagnetic potentials. The generalized electric potential is defined by $V = e \Phi - m \Phi_{\rm g}$. $V_{\rm int}$ includes the electron-electron potential due to Cooper-pairing of the electrons. It was shown in \cite{Schiff,Dessler} that in the presence of a gravitational field, the so-called Schiff--Barnhill field ${\bf K} = -\nabla V = e {\bf E} + m \nabla \Phi_{\rm g}$, and not ${\bf E}$ alone, vanishes in an ordinary conductor. DeWitt argues that if one now applies the BCS formalism to the Hamiltonian (\ref{68}), the result is that a modified Meissner effect is valid for $e {\bf B} - 2 m {\bf B}_{\rm g}$ instead of for ${\bf B}$ alone, where ${\bf B}=\nabla \times {\bf A}$ and ${\bf B}_{\rm g}=\nabla \times {\bf A}_{\rm g}$. The same is true for the flux quantization condition, which now reads
\be
\oint\limits_S (e {\bf B} - 2m {\bf B}_{\rm g}) \cdot d{\bf S} = n \frac{h c}{2}\ .
\ee

If one considers the setup of DeWitt described above, then the condition 
\be
\oint\limits_S (e {\bf B} - 2m {\bf B}_{\rm g}) \cdot d{\bf S} = 0
\ee
holds for the superconducting ring, in case that no external fields are applied and the mass in the centre is at rest. As soon as the mass inside the ring is rotated, it produces a Thirring--Lense field which acts on the superconductor. Since not ${\mathbf B}$ but the combination $e {\mathbf B} - 2m {\mathbf B}_{\rm g}$ vanishes, this means that a gravitomagnetic flux induces an electric current in the superconducting ring which could (theoretically) be measured. Using the formula for the Thirring--Lense effect, he makes an estimate of this current (here, SI units are used),
\be
I \sim \eps_0 c^4 \kappa m M v / e d,
\ee
where $\kappa = 8 \pi G/c^4$, $m$ is the electron mass, and $M$, $v$, and $d$ are the mass, rim velocity, and diameter of the rotating body, respectively. Taking the values \mbox{$M = 1$ kg} and $v = 300 \times 2 \pi \times d[{\rm m}]/2$ m/s, one obtains $I \approx 10^{-29}$ A.

Today, SQUID-type measurements are able to measure currents in the order of fA with relative uncertainties in the order of $10^{-3}$ \cite{Melcher,Feltin}. Thus these currents are presently out of experimental reach. It should also be noted that other effects, such as Thirring--Lense fields originating from the Earth, need to be considered in order to isolate the effects caused by the rotating mass.

\subsubsection{Currently proposed experiments}

The idea of Tajmar and de Matos \cite{Tajmar2,Tajmar3} is the combination of two effects: the consideration of the London moment as well as the gravitomagnetic field caused by the rotation of a superconducting mass. To this end, they consider a superconductor rotated about its axis and include gravitational drag effects. Following DeWitt \cite{DeWitt} and Ross \cite{Ross}, they alter the quantization conditions to include gravitomagnetic potentials as well. This yields modified London and Ginzburg--Landau quantization conditions. Instead of
(\ref{quantCond}), one now finds\footnote{The additional factor 2 (or 2/c) which appears in the following equations when comparing to the results of \cite{Tajmar2} arises because of the different definitions of the gravitoelectromagnetic potentials, see Sect. \ref{gravito}.}
\be
\oint\limits_C (m {\bf v}_{\rm s} + \frac{e}{c} {\bf A} - \frac{2m}{c} {\bf A}_{\rm g}) \cdot \D{\bf l} = n \frac{h}{2}\ ,
\ee
while instead of (\ref{additional}) one has
\be
\oint\limits_{S} {\bf B} \cdot \D{\bf S} - \frac{2m}{e} \oint\limits_{S} {\bf B}_{\rm g} \cdot \D{\bf S} + \frac{m c}{e^2 n_{\rm s}} \oint\limits_C {\bf j} \cdot \D{\bf l} + \frac{2 m c}{e} \mbox{\boldmath$\omega$} \cdot {\bf S}= n \frac{h c}{2 e}\ .
\ee
The inclusion of the gravitomagnetic field should in principle be witnessed in the measurements of the London moment as well as the Cooper pair mass. In analogy to the argumentation used without considering gravitomagnetic fields, Tajmar and de Matos consider the cases of a thick and thin superconducting ring,
respectively. For the thick ring, one obtains a modification of the London moment,
\begin{equation}
{\bf B} = -\frac{2 m c}{e} \mbox{\boldmath$\omega$} + \frac{2m}{e} {\bf B}_{\rm g\ },\label{LonMomentMod}
\end{equation}
and for the thin ring, a modified equation for the Cooper pair mass $m' = 2 m$ yields
\begin{equation}
\frac{h}{m'} = 2 S \left(\Delta \omega + \frac{\Delta B_{\rm g, lattice}}{c}\right)\ ,
\end{equation}
where the additional term arises from the contribution of the superconductor's neutral lattice structure.\\

According to the classical coupling scheme discussed in Sect. \ref{classical}, the gravitomagnetic field would be too small to be detectable. Considering the modified London moment (\ref{LonMomentMod}) and the coupling equation (\ref{coupling4}), one obtains
\be
\frac{B}{c} \approx \frac{2 m}{e} \mbox{$\omega$}\ , \qquad \frac{ B_{\rm g}}{c} \approx G \left(\frac{2 m}{e}\right)^2 \mbox{$\omega$} \ll \frac{B}{c}\ .
\ee
This would also imply negligible effects in the measurement of the Cooper pair mass. On the other hand, Tajmar and de Matos \cite{Tajmar2} suggest that this classical coupling between the two fields might be altered in the interaction of gravitational fields with quantum systems (cf. Sects. \ref{quant1}, \ref{quant2}), as is the case here. It is argued that the gravitomagnetic field could be responsible for the experimental discrepancies of the measurements of the London moment and the superconducting mass \cite{Hildebrandt,Tate}, which are correct only up to a few percent.

Assuming that the experimental discrepancies can be fully attributed to the inclusion of gravitomagnetic effects, Tajmar and de Matos find values of $B_{\rm g} \approx 7,000~c/2$ Hz for the measurement of the London moment by Hildebrandt \cite{Hildebrandt} and $B_{\rm g} \approx 1.65 \times 10^{-5}~c/2$ Hz for the determination of the Cooper pair mass by Tate \cite{Tate}.\footnote{Again, the values differ from those found in \cite{Tajmar2} due to the different definitions for the gravitoelectromagnetic potentials.} While these large values do not disqualify the inclusion of the gravitomagnetic effects in the calculations, since the presence of these fields is theoretically expected, it is strongly questionable whether these effects, which are expected to be much smaller than the above values, can be isolated from other systematic errors which arise in these experiments. Furthermore, the calculations in Tate's experiment are very sensitive to the effective mass of the Cooper pairs in Niobium, and thus already the correction of the theoretical mass would lead to better agreement with the experiment.

Finally, we want to mention that the same applies to the experiment conducted on rotating superfluids in \cite{Simmonds}. Tajmar and de Matos suggest extending the Josephson current which is measured to include the gravitomagnetic field ${\bf B}_{\rm g}$ \cite{Tajmar2}; however, the field which would be needed in order to strongly influence the experiment is very large. Again, there is no reason why it should be so based on the present knowledge about the coupling strength between quantum fluids and gravitational fields.

\subsection{Conclusion on proposed experiments with rotating superconductors}

To summarize, we have reviewed the approaches suggested in dealing with rotating superconductors in which weak gravitational fields have been taken into account. An estimate of DeWitt has shown that the induced current in a superconducting ring exposed to Thirring--Lense fields is out of experimental reach today, even if it is possible to isolate the effect caused by the rotating mass from those of the environment. Generally, it can be argued that there is not any theoretical or experimental knowledge as yet that would suggest performing further experiments on rotating superconductors, since any realistic estimates would not explain the proposed experimental discrepancies, for example in Hildebrandt's and Tate's experiments. It would thus be advisable, as already mentioned in Sect. \ref{concl1}, to first deepen the theoretical knowledge of the coupling between electro- and gravitoelectromagnetic fields.

\section{Summary and conclusion}

In this study, we have discussed possible interactions between two basic foundations of physics: quantum theory on the one hand and the theory of general relativity on the other. The basic ideas and theories were presented and important experiments as well as current propositions mentioned.

Since much research has recently concerned the interaction of gravitational fields with rotating and stationary quantum fluids (quantum systems with macroscopic phase coherence), we have discussed the proposed ideas and evaluated their results. Since gravity couples to all forms of energy, it is clear that coupling occurs, and we want to stress that it is of utmost importance to investigate this interaction further; however, we do not feel that the propositions lead to effects which can presently be measured in the laboratory.\\

To be more specific, the ideas regarding the generation (and detection) of gravitational waves via quantum systems should be considered theoretically in order to gain a better understanding of the exact coupling strength; only in this way, it is possible to assess the outcome of experiments conducted to this end. To our knowledge, there exists no general non-relativistic derivation of the coupling of an electron to both eletromagnetic and gravitational fields, neither in the post-Newtonian approximation nor in the background of a gravitational wave. Thus, it is not possible to make predictions about the coupling strengths. It is highly recommended that this derivation be performed in order to gain qualitative as well as quantitative results.

Summarizing the propositions for experiments using rotating superconductors, there does not seem to be evidence for the inclusion of the gravitomagnetic fields on experimental grounds. The classical coupling scheme discussed yields interactions which can be neglected. Of course, as a result of GR, the gravitoelectromagnetic fields should be considered in the calculations. Yet, as stated before, we recommend that, even though these effects are interesting in their theoretical origin and should be investigated further, this should be done in a theoretical context to arrive at quantum coupling mechanisms which could lead to experimental results.

\section{Acknowledgements}

This work was supported by the European Space Agency (ESA) under the Ariadna scheme of the Advanced Concepts Team, contract 18152/04/NL/MV. Our thanks go to the ESA, and especially to Andreas Rathke and Clovis de Matos for their comments on this article. We would also like to thank Alexander Altland, Friedrich Hehl, Siddhartha Lal, Bahram Mashhoon, Tobias Micklitz, Thomas Nattermann, Yuri Obukhov, and Bernd Rosenow for helpful discussions and J\"urgen Melcher, Gerd-Dietmar Willenberg, and Alexander Zorin from the Physikalisch-Technische Bundesanstalt for their expertise on current measurements.

\end{document}